\documentclass[a4paper,twocolumn,11pt,accepted=2022-04-28]{quantumarticle}

\pdfoutput=1
\usepackage[utf8]{inputenc}
\usepackage[english]{babel}
\usepackage[LGR,T1]{fontenc}
\usepackage{amsmath}
\usepackage{float}
\usepackage{graphicx}
\usepackage{amsmath,etoolbox}
\usepackage{amssymb}
\usepackage{xcolor}
\usepackage{hyperref}
\usepackage{soul}
\usepackage{bbm}
\usepackage{commath}
\usepackage{enumerate}
\usepackage{mathtools}
\usepackage{calrsfs}
\DeclareMathAlphabet{\pazocal}{OMS}{zplm}{m}{n}
\usepackage{upgreek}
\usepackage{calc}
\usepackage{accents}
\usepackage{siunitx}

\usepackage[numbers,sort&compress]{natbib}
\graphicspath{{fig/}} 

\newcommand{\declarebsfgreek}[2]{%
  \protected\csdef{bsf#1}{\mathord{\text{\bsfgreekfont#2}}}%
}
\newcommand{\bsfgreekfont}{\usefont{LGR}{cmss}{bx}{it}}
\declarebsfgreek{alpha}{a}
\declarebsfgreek{beta}{b}
\declarebsfgreek{gamma}{g}
\declarebsfgreek{delta}{d}
\declarebsfgreek{epsilon}{e}
\declarebsfgreek{zeta}{z}
\declarebsfgreek{eta}{h}
\declarebsfgreek{theta}{j}
\declarebsfgreek{iota}{i}
\declarebsfgreek{kappa}{k}
\declarebsfgreek{lambda}{l}
\declarebsfgreek{mu}{m}
\declarebsfgreek{nu}{n}
\declarebsfgreek{xi}{x}
\declarebsfgreek{omicron}{o}
\declarebsfgreek{pi}{p}
\declarebsfgreek{rho}{r}
\declarebsfgreek{sigma}{s}
\declarebsfgreek{tau}{t}
\declarebsfgreek{upsilon}{u}
\declarebsfgreek{phi}{f}
\declarebsfgreek{chi}{q}
\declarebsfgreek{psi}{u}
\declarebsfgreek{omega}{w}

\DeclareMathAlphabet{\pazocal}{OMS}{zplm}{m}{n}

\newcommand{\comments}[1]{}
\newcommand{\bea}{\begin{eqnarray}}
\newcommand{\eea}{\end{eqnarray}}

\makeatletter
\DeclareFontFamily{OMX}{MnSymbolE}{}
\DeclareSymbolFont{MnLargeSymbols}{OMX}{MnSymbolE}{m}{n}
\SetSymbolFont{MnLargeSymbols}{bold}{OMX}{MnSymbolE}{b}{n}
\DeclareFontShape{OMX}{MnSymbolE}{m}{n}{
	<-6>  MnSymbolE5
	<6-7>  MnSymbolE6
	<7-8>  MnSymbolE7
	<8-9>  MnSymbolE8
	<9-10> MnSymbolE9
	<10-12> MnSymbolE10
	<12->   MnSymbolE12
}{}
\DeclareFontShape{OMX}{MnSymbolE}{b}{n}{
	<-6>  MnSymbolE-Bold5
	<6-7>  MnSymbolE-Bold6
	<7-8>  MnSymbolE-Bold7
	<8-9>  MnSymbolE-Bold8
	<9-10> MnSymbolE-Bold9
	<10-12> MnSymbolE-Bold10
	<12->   MnSymbolE-Bold12
}{}

\newcommand{\ignore}[1]{}
\newcommand{\nobibentry}[1]{{\let\nocite\ignore\bibentry{#1}}}

\newcommand{\im}[1]{\text{Im}\,#1}

\begin{document}

\title{Bending the rules of low-temperature thermometry with periodic driving}


\author{Jonas Glatthard}
\affiliation{Department of Physics and Astronomy, University of Exeter, Exeter EX4 4QL, United Kingdom}
\email{J.Glatthard@exeter.ac.uk}

\author{Luis A. Correa}
\affiliation{Department of Physics and Astronomy, University of Exeter, Exeter EX4 4QL, United Kingdom}

\begin{abstract}
There exist severe limitations on the accuracy of low-temperature thermometry, which poses a major challenge for future quantum-technological applications. Low-temperature sensitivity might be manipulated by tailoring the interactions between probe and sample. Unfortunately, the tunability of these interactions is usually very restricted. Here, we focus on a more practical solution to boost thermometric precision---driving the probe. Specifically, we solve for the limit cycle of a periodically modulated linear probe in an equilibrium sample. We treat the probe--sample interactions \textit{exactly} and hence, our results are valid for arbitrarily low temperatures $ T $ and any spectral density. We find that weak near-resonant modulation strongly enhances the signal-to-noise ratio of low-temperature measurements, while causing minimal back action on the sample. Furthermore, we show that near-resonant driving changes the power law that governs thermal sensitivity over a broad range of temperatures, thus `bending' the fundamental precision limits and enabling more sensitive low-temperature thermometry. We then focus on a concrete example---impurity thermometry in an atomic condensate. We demonstrate that periodic driving allows for a sensitivity improvement of several orders of magnitude in sub-nanokelvin temperature estimates drawn from the density profile of the impurity atoms. We thus provide a feasible upgrade that can be easily integrated into low-$T$ thermometry experiments.
\end{abstract}



\maketitle

\section{Introduction}\label{sec:introduction}

Recent experimental progress is making it possible to cool systems down to unprecedented low temperatures. For instance, cold atomic gases have been pushed into the sub-nanokelvin regime, and even down to few picokelvins \cite{leanhardt2003cooling,bloch2012quantum,olf2015thermometry,deppner2021collective}. Yet \textit{measuring} such record-breaking temperatures accurately remains remarkably difficult \cite{olf2015thermometry}. This is not only due to technical issues, nor specific to cold atomic gases. There are fundamental physical limitations that render low-temperature thermometry inefficient \cite{hovhannisyan2018termometry_many-body}. Furthermore, at ultracold temperatures, infinitesimal-coupling treatments that rely on local thermalisation of probe systems become inadequate, since probe and sample build quantum correlations which can push their marginals far from the Gibbs state \cite{de2016local,correa2017thermometry_strongcoupling,miller2018energy}.

These shortcomings have motivated the stand-alone theoretical study of thermometric precision in the quantum regime, termed \textit{quantum thermometry} (see \cite{depasquale2019thermometry,mehboudi2019review} and references therein). In particular, results on the scaling laws that govern thermal sensitivity as the temperature $ T\rightarrow 0 $ \cite{hovhannisyan2018termometry_many-body,potts2019thermometry_fundamental,jorgensen2020tight} paint a grim picture. The signal-to-noise ratio of estimates on the temperature of gapped samples is exponentially suppressed as $ T $ decreases \cite{hovhannisyan2018termometry_many-body}, while in gapless systems it decays as a power law in $ T $ instead\footnote{A temperature-independent signal-to-noise ratio is possible in principle \cite{hovhannisyan2018termometry_many-body,mukherjee2019thermometry_control,henao2021thermometric}, but it requires engineering the probe--sample Hamiltonian, working with limiting-case spectral densities, or using adaptive strategies.} \cite{hovhannisyan2018termometry_many-body,potts2019thermometry_fundamental,jorgensen2020tight}. It is therefore essential to devise protocols that can enhance low-temperature sensitivity as much as possible. 

Engineering the probe--sample interactions \cite{correa2017thermometry_strongcoupling,mehboudi2019bosepolarons} can greatly improve thermometric precision, but the degree of control over the spectral properties of the sample is arguably limited. Alternatively, Mukherjee \textit{et al.} \cite{mukherjee2019thermometry_control} proposed addressing the probe with a periodic driving field. Specifically, it was shown that periodic driving allows to open new dissipation channels at tuneable frequencies and to tailor the weight of their contribution to the overall dissipative dynamics, ultimately yielding improved thermometric bounds. Periodic driving has also proven useful in other metrological settings \cite{mishra2021integrable,PhysRevLett.127.080504}.

Technically, the analysis in Ref.~\cite{mukherjee2019thermometry_control} combines weak-coupling Gorini--Kossakowski--Lindblad--Sudarshan (GKLS) quantum master equations \cite{davies1974markovian,gorini1976completely,lindblad1976generators} with Floquet theory \cite{PhysRevA.73.052311,alicki2012periodically}. Imposing the GKLS form in a master equation guarantees formal thermodynamic consistency and, for vanishingly weak dissipation, they may yield an excellent approximation to the steady state of a probe coupled to an ultracold sample. However, microscopically derived Born--Markov secular GKLS master equations are known to break down at any finite dissipation strength, as one moves to the $ T\rightarrow 0 $ limit \cite{correa2017thermometry_strongcoupling,mehboudi2019bosepolarons}. Namely, GKLS equations invariably predict local thermalisation of any probe at the sample temperature in the absence of driving. At ultracold temperatures, however, these do build strong correlations, which results in non-thermal marginals. Furthermore, using GKLS equations on open systems with tailored spectral densities may be questioned from a microscopic-derivation viewpoint whenever the resulting correlation functions in the sample are long lived \cite{breuer2002theory}. 

Here we study the impact of periodic driving on low-temperature thermometric precision, circumventing all these issues. We exploit the exact techniques from \cite{zerbe1995parametric,freitas2014linear_analytic,freitas2017linear_refrigerator} and solve for the non-equilibrium state of a periodically driven Brownian particle in a harmonic potential and dissipatively coupled to a linear equilibrium sample. Importantly, our results do not rely on any assumptions on the strength of the dissipation, the spectral structure of the sample, or its temperature. We provide explicit formulas for the limit-cycle state of the probe, exact up to any order in the strength of the drive, and use them to calculate the theoretical upper bound on the signal-to-noise ratio of temperature estimates drawn from arbitrary measurements on the probe \cite{mehboudi2019review}. Ultimately, we analyse the low-$ T $ scaling of the precision of such estimates. 

Firstly, we find that our calculations qualitatively agree with the findings in Ref.~\cite{mukherjee2019thermometry_control}; that is, the largest precision enhancements occur when the external drive is weak and near-resonant. For our model, thermometric precision without driving is known to obey a power law, with an exponent dictated by the low-frequency behaviour (Ohmicity) of the spectral density of the sample \cite{hovhannisyan2018termometry_many-body}. Interestingly, we find that near-resonant periodic driving effectively modifies this scaling over a broad range of temperatures, resulting in a large improvement of low-$T$ thermometry. We also demonstrate that, in order to benefit from this precision boost, it is not necessary to synchronise the measurements with the driving field---the advantage survives time-averaging of the oscillating asymptotic probe state. 

In order to illustrate the practical potential of this technique on a concrete platform, we considered impurity thermometry in a Bose--Einstein condensate. We exploit the fact that tightly confined atomic impurities immersed in a cold atomic gas with a large condensed fraction can be modelled with our Brownian-motion model \cite{lampo2017bosepolarons,lampo2018bosepolarons,mehboudi2019bosepolarons}. Instead of evaluating the precision limits, we adopt a more applied perspective. Working with the asymptotic time-averaged state of the probe, we evaluate the responsiveness of the density profile of the impurity cloud to small temperature fluctuations. For typical experimental parameters, we show that near-resonant driving leads to a sensitivity enhancement of several orders of magnitude at temperatures $ T \lesssim\SI{1}{\nano\kelvin} $.  

Finally, we take into account a side-effect of addressing the probe with an external field---the heating of the sample. Specifically, we calculate the coarse grained long-time probe-sample heat flow. We find that, even when driving near resonance, the back-action on the sample is a fourth-order effect in the strength of the external drive. 

This paper is structured as follows: The model is introduced in Sec.~\ref{sec:hamiltonian} and an outline of the main steps in the calculation of its non-equilibrium limit-cycle state is given in Sec.~\ref{sec:cycle} (the full calculation is in Appendix~\ref{app:solving}). After briefly introducing the necessary estimation-theory tools in Sec.~\ref{sec:thermometry}, we present our main results in Sec.~\ref{sec:main-result}, and apply them to impurity thermometry in Sec.~\ref{sec:applications}. Sample heating is discussed in Sec.~\ref{sec:heating}, with technical details deferred to Appendices \ref{app:current} and \ref{app:mathieu}. Finally, in Sec.~\ref{sec:conclusions} we summarise and draw our conclusions.

\section{The model and its limit cycle}\label{sec:model}

\subsection{Hamiltonian}\label{sec:hamiltonian}

Let the Hamiltonian of our probe be
\begin{equation}
	 \pmb H_S(t) = \frac{1}{2} \omega^2(t)\,\pmb x^2 + \frac{1}{2}\,\pmb p^2. \label{eq:system_Hamiltonian}
\end{equation}
In all what follows, boldface symbols denote operators and we shall work in units of $m=\hbar=k_B=1$ unless stated otherwise. We choose a time-dependent frequency of the form
\begin{equation}
	\omega(t)^2 = \omega_0^2 + \upsilon \sin{\omega_d t} \label{eq:driving_frequency}.
\end{equation}
The role of the sample will be played by a linear heat bath with Hamiltonian
\begin{equation}\label{eq:bath_hamiltonian}
	\pmb H_B = \sum\nolimits_\mu \frac{1}{2} \omega_\mu^2 m_\mu \pmb x_\mu^2  + \frac{1}{2 m_\mu} \pmb p_\mu^2,
\end{equation}
while the dissipative coupling between probe and sample is
\begin{equation}\label{eq:sys-bath}
	\pmb H_I = \pmb x\,\sum\nolimits_\mu g_\mu \pmb x_\mu.
\end{equation}
The interaction constants $g_\mu$ shape the spectral density
\begin{equation*}
	J(\omega) \coloneqq \pi \sum_\mu \frac{g_\mu^2}{2 m_\mu \omega_\mu} \delta(\omega - \omega_\mu),
\end{equation*}
which we take to be Ohmic with algebraic cutoff
\begin{equation}
	J(\omega) = \frac{2 \gamma\,\omega}{1+(\omega/\omega_c)^2} \label{eq:spectral_density}.
\end{equation}
Here $\omega_c$ stands for the cutoff frequency of the bath. We refer such spectrum as Ohmic since $ J(\omega)\sim \omega^s $ at low frequency with $ s=1 $. In contrast, $ s>1 $ and $ s<1 $ would correspond to super- and sub-Ohmic spectral densities, respectively.

The exact Heisenberg equations of motion for the system can be written in compact form as (cf. Appendix~\ref{app:eom}) 
\begin{multline}
\ddot{\pmb x}(t)+\big(\omega^2(t)+\omega_R^2\big)\,\pmb x(t) \\ - \int_0^t  \chi(t-\tau)\, \pmb x(\tau) d \tau = \pmb F(t), \label{eq:QLE_driven}
\end{multline}
where we assume that the bath's marginal is initially a state of thermal equilibrium. At this point, the initial state of the probe can be fully general, though we will later restrict ourselves to Gaussian initial states. The dissipation kernel $\chi(t)$ in Eq.~\eqref{eq:QLE_driven} is
\begin{equation}
	\chi(t)=\frac{2}{\pi} \Theta(t)\int_0^\infty J(\omega) \sin{(\omega t)}\,d \omega \label{eq:dissipation_kernel},
\end{equation}
and the stochastic force $\pmb F(t)$ is defined in Appendix~\ref{app:eom}. $\Theta(t)$ stands for the Heaviside step function. The frequency shift $\omega_R$ in Eq.~\eqref{eq:QLE_driven} is
\begin{equation}
\omega_R^2 \coloneqq \frac{2}{\pi} \int_0^\infty \frac{J(\omega)}{\omega} d \omega \label{eq:counter_term}.
\end{equation}
This is introduced in the Hamiltonian to compensate for the distortion on the probe's potential due to its coupling to the bath \cite{weiss1999}. For our $J(\omega)$ in Eq.~\eqref{eq:spectral_density}, it evaluates to $ \omega_R^2 = 2\,\gamma\,\omega_c $.

\subsection{Limit cycle}\label{sec:cycle}

Let $g(t,t')$ be the Green's function for Eq.~\eqref{eq:QLE_driven}, i.e.,
\begin{multline}\label{eq:greens-text}
    \partial_t^2 g(t,t') + \big(\omega^2(t)+\omega_R^2\big)\, g(t,t') \\- \int_0^t \chi(t-\tau)\,g(\tau,t')\,d\tau = \delta(t-t').
\end{multline}
If a stable limit cycle exists (cf. Appendix~\ref{app:mathieu}) $ g(t,t') $ can be cast as
\begin{equation}
	g(t,t')=\frac{1}{2 \pi} \sum_{k=-\infty}^\infty \int_{-\infty}^\infty  a_k(\omega) \mathrm{e}^{i \omega (t-t')} \mathrm{e}^{i k \omega_d t} d \omega  \label{eq:Green-function-and-As}.
\end{equation}
The detailed derivation of \eqref{eq:Green-function-and-As} is deferred to Appendix~\ref{app:covcycle} (see also Ref.~\cite{freitas2017linear_refrigerator}). As we shall now see, knowledge of the amplitudes $ a_k(\omega) $ allows us to fully characterise the limit cycle of the probe. 

To proceed further, let us insert \eqref{eq:Green-function-and-As} into \eqref{eq:greens-text}, which leads to the following relationship between the amplitudes $a_k(\omega)$ 
\begin{multline}
	 \hat{g}_0[i (\omega + k \omega_d)]^{-1}a_k(\omega) + \\ \sum\nolimits_{l \neq 0}\,b_l\,a_{k-l} (\omega) = \delta_{k0}, \label{eq:A}
\end{multline}
where $\delta_{ij}$ is a Kronecker delta and the periodic driving has been Fourier expanded as
\begin{equation}\label{eq:frequency-Fourier}
    \omega(t)^2 = \omega_0^2 + \sum\nolimits_{l\neq 0 } b_l\,\mathrm{e}^{i\,l\,\omega_d\,t}.
\end{equation}
The notation $\hat{g}_0(s)$ in Eq.~\eqref{eq:A} stands for
\begin{equation}\label{eq:g0}
	\hat{g}_0(s) \coloneqq \big[s^2 + \omega_0^2 + \omega_R^2 - \hat{\chi}(s)\big]^{-1},
\end{equation}
where $ \hat{\chi}(s) $ is the Laplace transform of the dissipation kernel, i.e., $ \hat{\chi}(s) = \int_0^\infty dt\, \mathrm{e}^{-st}\, \chi(t) $. We shall implicitly work with $\hat{f}(i \omega)=\hat{f}(i \omega + 0^+)$. The choice of notation in \eqref{eq:g0} is due to the fact that $ \hat{g}_0(s) $ is the Laplace transform of the Green's function in Eq.~\eqref{eq:greens-text} in the undriven case. 

For our choice of spectral density we have
\begin{equation}
	\hat{\chi}(i \omega) = \frac{2 \gamma \omega_c^2}{\omega_c + i \omega}.
\end{equation}
In turn, since our driving in Eq.~\eqref{eq:driving_frequency} is sinusoidal the only non-vanishing coefficients in the expansion \eqref{eq:frequency-Fourier} are $b_{\pm 1} = \mp\,i\,\upsilon/2$.

One can solve Eq.~\eqref{eq:A} for $ a_k(\omega) $ self-consistently as
\begin{multline}\label{eq:general-self-consistent-ak}
	a_k^{(n)}(\omega) = \hat{g}_0[i (\omega + k \omega_d)] \\ \times\big( \delta_{k0} - \sum\nolimits_{l \neq 0} b_l\,a_{k-l}^{(n-1)}(\omega) \big),
\end{multline}
where $ a_k^{(n)}(\omega) $ is the solution after $ n $ iterations, starting from
\begin{equation*}
    a_k^{(0)}(\omega) = \hat{g}_0[i(\omega+k\omega_d)]\,\delta_{k0}.
\end{equation*}
Each iteration of Eq.~\eqref{eq:general-self-consistent-ak} adds only terms of higher order in $\upsilon$, and $a_k=\lim_{n \rightarrow \infty} a_k^{(n)}$.

Substituting \eqref{eq:frequency-Fourier} and \eqref{eq:general-self-consistent-ak} into \eqref{eq:Green-function-and-As} yields the Green's function $ g(t,t') $ of the equation of motion of the system \eqref{eq:QLE_driven} in the limit cycle. In fact, we can explicitly write the (time-dependent) asymptotic state of the system in terms of the Fourier coefficients $ a_k(\omega) $ and $ b_k $. To that end, we exploit the fact that the Hamiltonian is quadratic in positions and momenta, thus generating a \textit{Gaussianity}-preserving dynamics \cite{ferraro2005gaussian}. That is, given a Gaussian initial condition with vanishing first moments, knowledge of the covariances $ \sigma_{xx} = \langle \pmb{x}^2 \rangle $, $ \sigma_{xp} = \frac12\langle \pmb{x}\pmb{p} + \pmb{p}\pmb{x} \rangle = \frac12\langle \{ \pmb x,\pmb p \} \rangle$ and $ \sigma_{pp} = \langle \pmb{p}^2 \rangle $ is enough to fully characterise the state of the probe. 

Specifically, the limit-cycle covariances of the probe in contact with an equilibrium bath are
\begin{subequations}\label{eq:covariances-limit-cycle}
\begin{align}
	\sigma_{\alpha\alpha}^\text{lc}(t) &= \mathrm{Re}\,\sum\nolimits_{j\,k}\varsigma_{jk}^{\alpha\alpha}\, \mathrm{e}^{i\, \omega_d (j-k)\,t},  \label{eq:sxxt}\\
	\sigma_{xp}^\text{lc}(t) &= \mathrm{Im}\,\sum\nolimits_{j\,k}\varsigma_{jk}^{xp}\, \mathrm{e}^{i\, \omega_d (j-k)\,t},  \label{eq:sxpt}
\end{align}
\end{subequations}
where $ \alpha=\{x,p\} $ and the super-index `lc' indicates `limit cycle'. The coefficients $ \varsigma^{\alpha\beta}_{jk} $ are 
\begin{subequations}\label{eq:covariances_coefficients-limit-cycle}
\begin{align}
	\varsigma_{jk}^{xx} &=\frac{1}{2}\int_0^\infty  a_j(\omega )\,\hat{\mu}(\omega)\,a_k^*(\omega)\,d\omega, \label{sxx} \\
 \varsigma_{jk}^{pp} &=\frac{1}{2}\int_0^\infty (\omega + j\, \omega_d)\,(\omega + k\,\omega_d)\nonumber\\
 & \qquad\qquad\qquad \times a_j(\omega)\,\hat{\mu}(\omega)\, a_k^*(\omega)\,d\omega, \label{spp} \\
 \varsigma_{jk}^{xp}&=\frac{1}{2}\int_0^\infty (\omega + k\,\omega_d)\, a_j(\omega)\hat{\mu}(\omega)\,a_k^*(\omega)\,d\omega. \label{sxp}
\end{align}
\end{subequations}
Above, we have introduced the notation
\begin{equation}
	\hat{\mu}(\omega) \coloneqq \frac{2}{\pi}\,J(\omega)\, \coth{\left(\omega/2 T\right)}, \label{eq:nk}
\end{equation}
and the asterisk denotes complex conjugation. The full derivation can be found in Appendix~\ref{app:covcycle}. 

Note that, for our choice of driving, $ a_k $ is $ \pazocal{O}(\upsilon^k) $. Since $ a_k(\omega) = a_k^{(2)}(\omega) + \pazocal{O}(\upsilon^3) $, we may stop after the second iteration in Eq.~\eqref{eq:general-self-consistent-ak} for sufficiently small $ \upsilon $ (i.e., weak periodic modulation). Explicitly, this gives us
\begin{subequations}\label{eq:a_k-2nd-order}
\begin{multline}
	a_0(\omega) \simeq \hat{g}_0(i \omega) \\+ \sum_{l \neq 0} \hat{g}_0(i \omega)\,b_l\, \hat{g}_0[i (\omega - l \omega_d)]\,b_{-l}\,\hat{g}_0(i \omega).
\end{multline}
\vspace{-0.7cm}
\begin{multline}
	a_{k\neq 0}(\omega) \simeq -\hat{g}_0[i (\omega + k \omega_d)]\,b_k\,\hat{g}_0(i \omega) \\ 
	+\sum\nolimits_{l \neq \{0, k\}} \hat{g}_0[i (\omega + k \omega_d)] b_l\\
	\times \hat{g}_0[i (\omega + (k-l) \omega_d)]\,b_{k-l}\,\hat{g}_0(i \omega),
\end{multline}
\end{subequations}
which facilitates calculations. For our simple driving, the only non-zero amplitudes at second order are thus $ a_0^{(2)}(\omega), a_{\pm 1}^{(2)}(\omega) $ and $ a_{\pm 2}^{(2)}(\omega) $. We emphasise, however, that Eqs.~\eqref{eq:covariances-limit-cycle} and \eqref{eq:covariances_coefficients-limit-cycle} are \textit{non-perturbative} in the strength $ \upsilon $ of the external driving, so that aribitrarily high orders can be considered. 

\section{Ultimate thermometric precision}\label{sec:thermometry}

The Cram\'er--Rao inequality \cite{cramer2016mathematical,rao1945information} places a lower bound the statistical uncertainty on any parameter inferred from a large number of measurements $ N $ by means of an unbiased estimator. This bound has been extensively used in high-precision sensing and, in particular, in low-temperature estimation (see \cite{mehboudi2019review,depasquale2019thermometry} and references therein). Namely, the uncertainty in a temperature estimate scales as
\begin{equation}\label{eq:cramer-rao}
	\delta T_ \mathrm{est} \geq \frac{1}{\sqrt{N\pazocal{F}(\pmb\varrho_T)}},
\end{equation}
where $\pazocal{F}(\pmb\varrho)$ is the `quantum Fisher information' calculated on the temperature-dependent state of the probe $ \pmb\varrho_T $ as
\begin{equation}
	\pazocal{F}(\pmb\varrho_T) = -2 \lim_{ \tau \rightarrow 0} \frac{\partial^2  \mathbb{F} (\pmb\varrho_T,\pmb\varrho_{T+\tau})}{\partial \tau^2}, \label{eq:QFI}
\end{equation}
and the Uhlmann fidelity $ \mathbb{F}(\pmb\varrho_1,\pmb\varrho_2) $ is defined as
\begin{equation*}
	\mathbb{F} (\pmb\varrho_1,\pmb\varrho_2) = \mathrm{tr}\left(\sqrt{\sqrt{\pmb\varrho_1}\,\pmb\varrho_2\, \sqrt{\pmb\varrho_1}}\right)^2.
\end{equation*}
While the Cram\'er--Rao bound is inadequate when working with scarce data \cite{rubio2021global}, it has proven useful to understand the fundamental scaling laws of low-temperature thermometry \cite{correa2017thermometry_strongcoupling,hovhannisyan2018termometry_many-body,potts2019thermometry_fundamental,henao2021thermometric}.

The quantum Fisher information in Eq.~\eqref{eq:QFI} is particularly simple to evaluate on (single-mode) Gaussian states. In that case, the fidelity between states $ \pmb\varrho_1 $ and $ \pmb\varrho_2 $ with covariance matrices $ \mathsf{\Sigma}_1 $ and $ \mathsf{\Sigma}_2 $, respectively, can be cast as \cite{Scutaru1998fidelity}
\begin{equation}\label{eq:fisher-explicit}
	\mathbb{F} (\mathsf\Sigma_1,\mathsf\Sigma_2) = 2 \left( \sqrt{\varkappa+1}-\sqrt{\lambda} \right)^{-1},
\end{equation}
where $\varkappa = 4 \det{\left(\Sigma_1+\Sigma_2\right)}$ and $\lambda =(4 \det \Sigma_1-1)(4 \det \Sigma_2-1)$. In the formula above we have taken into account that the first-order moments vanish in our case.

\begin{figure}[t]
\includegraphics[width=0.45\textwidth]{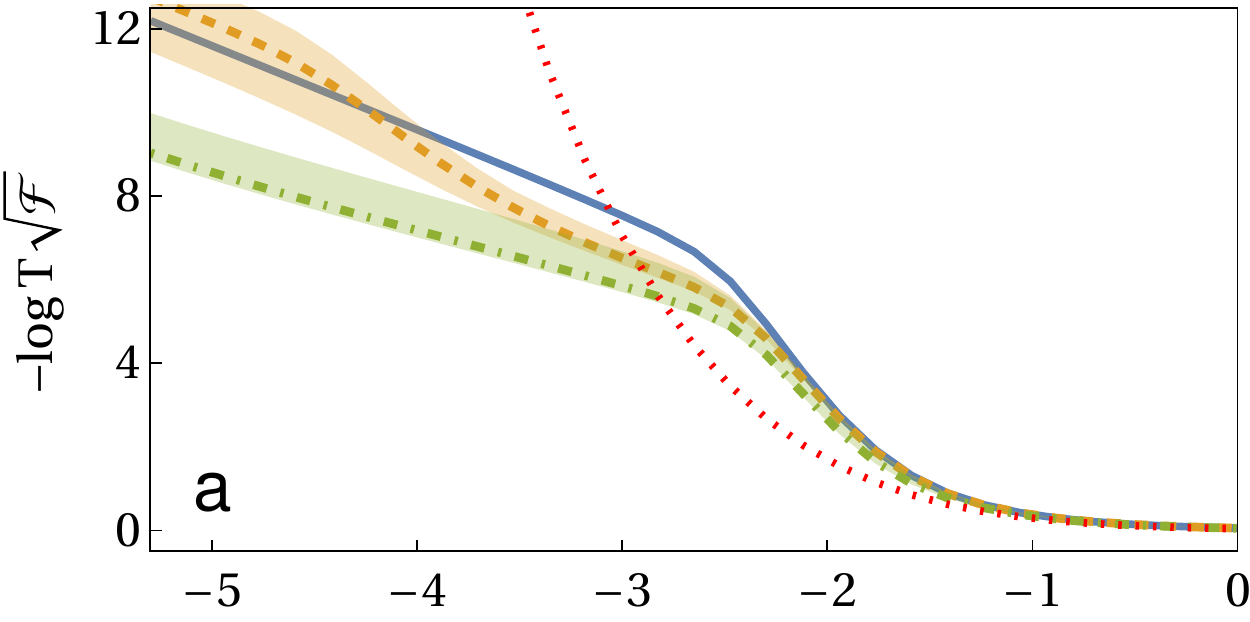}
\includegraphics[width=0.45\textwidth]{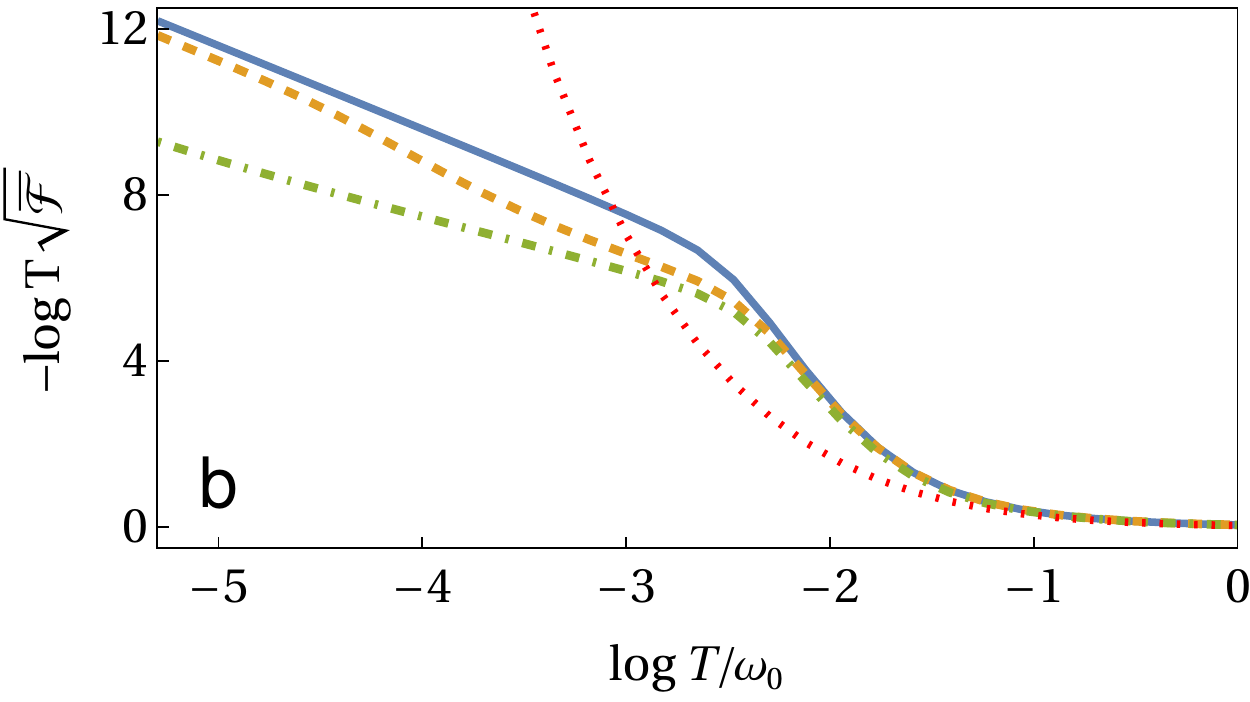}
\caption{\textbf{(a)} Instantaneous (inverse) thermal sensitivity in the limit cycle. Driving near resonance with $\omega_d = 0.9 \omega_0$ (dashed orange) and $\omega_d = 0.995 \omega_0$ (dot-dashed green) improves the thermal sensitivity of the probe. For comparison, the undriven case (solid blue) shows the expected $1/T^2$ scaling as $T \rightarrow 0$ \cite{hovhannisyan2018termometry_many-body}, while for a probe in a Gibbs state with respect to its local Hamiltonian (dotted red), the relative error grows exponentially. The limit cycle was sampled at 10 equispaced times which span the shaded regions around the curves. \textbf{(b)} Lower bound on the (inverse) thermal sensitivity from the time-averaged Fisher information $ \overline{\pazocal{F}(\pmb\varrho_T(t))} $. In both plots $\omega_0=1$, $\upsilon=0.1$, $\gamma=0.01$, and $\omega_c=100$.}
\label{fig1}
\centering
\end{figure}

\section{Results and discussion}

\subsection{Enhancement of low-$T$ sensitivity}\label{sec:main-result}

We are now in a position to assess the impact of periodic driving on the thermal sensitivity of the probe at arbitrarily low temperatures. To do so, we simply apply Eqs.~(\ref{eq:cramer-rao}--\ref{eq:fisher-explicit}) on the limit-cycle covariances from Eqs.~\eqref{eq:covariances-limit-cycle} and \eqref{eq:covariances_coefficients-limit-cycle}. It must be noted that, in contrast with master-equation methods, the parameter ranges that we may study is not limited by time-scale separation assumptions. That is, Eqs.~\eqref{eq:covariances-limit-cycle} and \eqref{eq:covariances_coefficients-limit-cycle} yield the \textit{exact} stationary state, only assuming Gaussianity for the initial condition, and that the sample is initially in thermal equilibrium. For our calculations below we do approximate the amplitudes $ a_k(\omega) $ to second and fourth order in $\upsilon$, which effectively limits our analysis to weak driving\footnote{More precisely, our covariances $\sigma_{\alpha\beta}^\text{lc}(t)$ are accurate to second (or fourth) order in $ \upsilon $. Care must be taken, however, since the covariance undergoes further manipulations in Eq.~\eqref{eq:fisher-explicit}, which generates higher-order terms which are not explicitly removed. Nonetheless, one can verify that the thermal sensitivity calculated from the second-order covariances does agree with the fourth-order calculations.}---but \textit{not} to weak dissipation or short-lived bath correlations. Furthermore, as we argue below, weak driving is a practical choice to minimise the heating of the sample (cf. \ref{app:current} and \ref{app:mathieu}).

We take the signal-to-noise ratio $ \frac{1}{\sqrt{N}}(T/\delta T_\text{est}) $ as the figure of merit, which we refer-to as \textit{thermal sensitivity}. The lower bound on the inverse of this quantity---as given by the Cram\'er--Rao inequality---is plotted in Fig.~\ref{fig1}(a) versus the temperature $ T $, for different modulations $ \omega_d $. The calculation is repeated considering various times within the limit cycle, since the asymptotic state of the probe still depends (periodically) on time. 

We observe that weak dissipation strength and near-resonant modulation ($\omega_d \approx \omega_0$) does boost sensitivity by several orders of magnitude in the low-temperature regime. More interestingly, the power law obeyed by the low-temperature sensitivity changes as a result of the driving. It is known that the Fisher information on temperature of a Brownian particle scales as $ \pazocal{F}(\pmb\varrho_T) \sim T^{2\,s} $ at $T\rightarrow 0$ \cite{hovhannisyan2018termometry_many-body} where $ s $ is the Ohmicity parameter introduced in Sec.~\ref{sec:hamiltonian}; in our case, $ s=1 $. In contrast, we find that for the power law is \textit{effectively} modified to $ \pazocal{F}(\pmb\varrho_T) \sim T $ as $ \omega_d\rightarrow\omega_0$. Hence, near-resonant periodic driving renders low-temperature thermometry much more efficient. This is our main result. 

By considering larger dissipation strengths we have also observed that maintaining a substantial advantage over the undriven case requires stronger driving $ \upsilon $. This is due to the fact that strong dissipation is capable of largely improving the low-temperature sensitivity \cite{correa2017thermometry_strongcoupling} on its own, which masks any improvements due to the drive. 

As already noted, the thermal sensitivity of the driven probe oscillates in time, spanning the shaded regions around the plotted curves in Fig.~\ref{fig1}(a). In order to get a clearer picture of the expected improvement in sensitivity, we may prefer to work with the time-averaged limit-cycle
\begin{equation}\label{eq:averaged-state}
\overline{\pmb{\varrho}}_T \coloneqq \frac{\omega_d}{2 \pi} \int_0^{2 \pi/\omega_d}\pmb\varrho_T(t)\,dt.
\end{equation}
Such average generally yields a non-Gaussian state lying on the convex hull of Gaussian states. As a result, we can no longer use Eq.~\eqref{eq:fisher-explicit} to evaluate the thermal sensitivity. Instead, we exploit of the extended convexity property of the quantum Fisher information \cite{Alipour2015convexity}, which becomes 
\begin{equation}\label{eq:QFI-time-averaged-state}
	\pazocal{F}_T \left[\int_0^{2\pi/\omega_d}\pmb\varrho_T(t)\,dt\right] \le  \int_0^{2\pi/\omega_d} \pazocal{F}_T[\pmb\varrho_T(t)]\,dt.
\end{equation}
Hence, by averaging the quantum Fisher information of the probe over the limit cycle, we can bound the instantaneous thermal sensitivity from below. This is illustrated in Fig.~\ref{fig1}(b), which suggests that the scaling improvement for near-resonant driving does survive time averaging. Indeed, in Sec.~\ref{sec:applications} we give, in a concrete example, a lower bound to the left hand side of Ineq. \eqref{eq:QFI-time-averaged-state} which also exhibits an enhanced low-temperature scaling from the drive. This has an important practical consequence: there would be no need to synchronise the measurements on the probe with the drive in order to benefit from the precision enhancement.

\subsection{Application: Impurity thermometry in a Bose--Einstein condensate}\label{sec:applications}

\begin{figure}[t]
\includegraphics[width=0.45\textwidth]{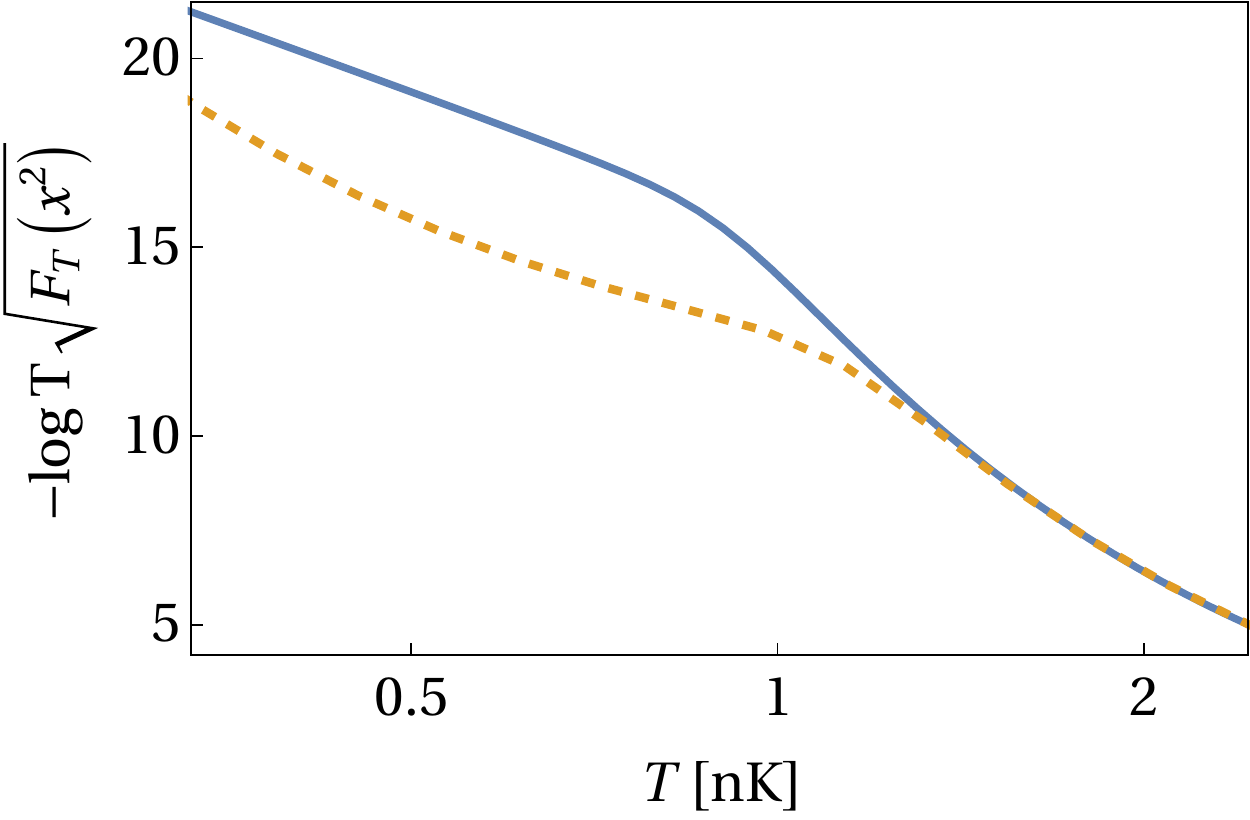}
\caption{Sensitivity improvement of the position quadrature of an impurity in a cold atomic gas due to periodic modulation of the trapping potential. The limit-cycle-averaged position dispersion of Yb impurities immersed in a BEC of K is used to estimated the temperature of the latter. The trapping frequency of for Yb is, on average, $\omega_I=2 \pi \times \SI{375}{\hertz}$, and is sinusoidally modulated with amplitude $\upsilon=0.2 \omega_I $ and frequency $\omega_d=0.8 \omega_I$. The resulting sensitivity (dashed orange) can be greatly enhanced with respect to the unmodulated case (solid blue). It is worth noting that in the unmodulated case, position dispersion is a quasi-optimal temperature indicator \cite{correa2017thermometry_strongcoupling}. The plotted temperature range of $\SI{0.5}{\nano\kelvin}$--$\SI{2}{\nano\kelvin}$ is within experimental reach \cite{leanhardt2003cooling,olf2015thermometry,aveline2020picokelvin}. The other parameters are $N_B=5000$, $\omega_B = 2 \pi \times \SI{750}{\hertz}$, $g_{IB}=0.55 \times 10^{39}\SI{}{\joule\metre}$ and $g_B=3 \times 10^{39}\SI{}{\joule\metre}$.}
\label{fig3}
\centering
\end{figure}

Thermometry in ultracold gases becomes particularly challenging as temperatures decrease. The conventional time-of-flight absorption imaging method can no longer be used on bosonic gases with a large condensed fraction, and alternatives must be sought \cite{olf2015thermometry}. Specifically, the temperature of a Bose--Einstein condensate (BEC) may be experimentally assessed using atomic impurities as probes \cite{PhysRevLett.109.235301,olf2015thermometry,PhysRevA.93.043607}. 
In the appropriate limits, a dilute gas of impurity atoms co-trapped alongside a BEC can be described with our simple quantum-Brownian-motion model \cite{lampo2017bosepolarons, lampo2018bosepolarons}. We may thus investigate whether periodic modulation of the confining potential of the impurities can enhance their thermal sensitivity at ultra-low temperatures. As we will see, it does.

Specifically, we consider an atomic impurity of mass $m_I$ in a 1D harmonic trap with frequency $\omega_I$. In turn, the co-trapped BEC is comprised of $N_B$ atoms of mass $m_B$ at temperature $T$, and is trapped in a harmonic potential with frequency $\omega_B$. One may show that, if the temperature is sufficiently low and the impurity is tightly confined around the centre of the condensate trap, its dynamics is effectively described by the model in Sec.~\ref{sec:hamiltonian}, with spectral density \cite{lampo2018bosepolarons}
\begin{equation}
	J(\omega) = 2\,\gamma_0\,\omega^4\, \Theta(\omega_B - \omega) \label{eq:super-Ohmic-spectral-density},
\end{equation}
where the constant $ \gamma_0 $ is given by
\begin{equation*}
    \gamma_0=\frac{\pi\,g_B}{\omega_B^4\,r^3}\left( \frac{g_{IB}\, \mu}{g_B\,\hbar\,\omega_B}\right)^2,
\end{equation*}
and $r$ and $\mu$ stand for the the Thomas--Fermi radius and the chemical potential, respectively
\begin{align*}
	r &= \sqrt{2 \mu/(m_B \omega_B^2)},\\
	\mu &= \left( \frac{3}{4 \sqrt{2}}\,g_B\,N_B\,\omega_B\,\sqrt{m_B}, \right)^{3/2}.
\end{align*}
Here, $g_{IB}$ is the interspecies (i.e., probe--sample) interactions, and $g_B$ stands for the interatomic interactions within the sample. $ J(\omega) $ also incorporates a hard cutoff through the Heaviside step function $\Theta(\omega_B - \omega)$. Notice that, within this section, we work in SI units. 

Being able to tackle this problem without quantum master equations is particularly important, since we are dealing with a super-Ohmic spectral density. Environments with such power spectra have longer-lived correlation functions than their Ohmic counterparts, which could easily compromise the Born--Markov approximation underpinning most master-equation analyses.

Since the resulting spectral density differs from our Eq.~\eqref{eq:spectral_density}, we must recalculate the Laplace transform $\hat{\chi}$ of the corresponding dissipation kernel. This reads \cite{lampo2018bosepolarons}
\begin{multline}
	\hat{\chi}(i\omega) =  \frac{\gamma_0\,\omega_B}{\pi} \\- \frac{2\,\gamma_0\,\omega^2}{\pi\,\omega_B^3} \left(\omega_B^2+\omega^2\, \log{\frac{\omega^2}{\omega^2+\omega_B^2}}\right).
\end{multline}
The only other difference with Sec.~\ref{sec:hamiltonian} is that the frequency shift $ \omega_R^2 $ may be entirely neglected in the regime of validity of the microscopic derivation of Eq.~\eqref{eq:super-Ohmic-spectral-density} \cite{mehboudi2019bosepolarons}. Otherwise, Eqs.~\eqref{eq:covariances-limit-cycle} and \eqref{eq:covariances_coefficients-limit-cycle} are directly applicable to this problem.

Rather than assuming a precise time control in the interrogation of the probe, we work with the time-averaged state from Eq.~\eqref{eq:averaged-state}, thus getting a better idea of the sensitivity gain that may be expected in practice. Also, instead of searching for ultimate precision bounds, we ask ourselves what is the \textit{responsiveness} of the position dispersion $ \pmb{x}^2 $ of the impurities to temperature fluctuations in the limit cycle. This is indeed an experimentally accessible temperature estimator \cite{PhysRevA.85.023623,PhysRevA.93.043607}, and its responsiveness may quantified as \cite{braunstein1994informationgeometry, toth2014metrology, mehboudi2019review}
\begin{equation}\label{eq:error-propagation}
F_T(\pmb x^2)\coloneqq\frac{\big\vert\partial_T \langle\pmb{x}^2\rangle_{\overline{\pmb\varrho}_T}\big\vert^2}{(\Delta \pmb{x}^2)^2}\leq \pazocal{F}_T,
\end{equation}
where the subindices emphasise that averages are taken over $ \overline{\pmb\varrho}_T $ and thus,
\begin{equation*}
    (\Delta\pmb{x}^2)^2 = \langle \pmb x^4 \rangle_{\overline{\pmb\varrho}_T} - (\langle \pmb x^2 \rangle_{\overline{\pmb\varrho}_T})^2.
\end{equation*}

Due to linearity, we can write $ \langle \pmb x^2 \rangle_{\overline{\pmb\varrho}_T} $ as $ \overline{\sigma_{xx}^\text{lc}(t)} $, where the bar indicates limit-cycle average. The temperature-dependence in Eqs.~\eqref{eq:covariances_coefficients-limit-cycle} is entirely contained in the noise kernel $ \hat{\mu}(\omega) $, which allows to readily calculate $ \partial_T\overline{\sigma_{xx}^\text{lc}} $. Furthermore, note that only those terms with $j=k$ from Eq.~\eqref{eq:covariances-limit-cycle} survive time averaging. On the other hand, to evaluate the denominator of Eq.~\eqref{eq:error-propagation} we use the fact that the instantaneous state of the probe is always Gaussian, so that the identity $ \langle \pmb x^4(t) \rangle = 3 \langle \pmb x(t)^2 \rangle^2$ holds at all times. Exploiting linearity again, we may write
\begin{align*}
    \langle \pmb x^4 \rangle_{\overline{\pmb\varrho}_T} - (\langle \pmb x^2 \rangle_{\overline{\pmb\varrho}_T})^2 = 3\overline{\sigma_{xx}^\text{lc}(t)^2}-(\,\overline{\sigma_{xx}^\text{lc}(t)}\,)^2.
\end{align*}

The temperature-dependence of the responsiveness $ F_T(\pmb x^2) $ is plotted in Fig.~\ref{fig3} with and without driving, for the cold atomic mixture studied in \cite{mehboudi2019bosepolarons}. As it can be seen, applying weak near-resonant driving to the confining potential of the impurities yields sensitivity improvements of several orders of magnitude at temperatures around $k_B T \sim \hbar (\omega_0-\omega_d)/4$. For typical parameters this corresponds to the $ \SI{0.1}{\nano\kelvin}$--$\SI{1}{\nano\kelvin}$ range, which is of experimental interest \cite{leanhardt2003cooling,olf2015thermometry,aveline2020picokelvin}, albeit extremely challenging for conventional thermometry \cite{olf2015thermometry}. Closer to resonance, we again observe improved scaling. 

Although we have chosen cold atoms for illustration, the same basic principle can be applied to other setups, like superconducting microresonators \cite{zmuidzinas2012resonators}. Switching to typical trapping frequencies in the \SI{}{\giga\hertz} range would result in improved temperature sensing at millikelvins, which is directly relevant for this platform.

\subsection{Heat dissipation}\label{sec:heating}

\begin{figure}[t!]
\includegraphics[width=0.45\textwidth]{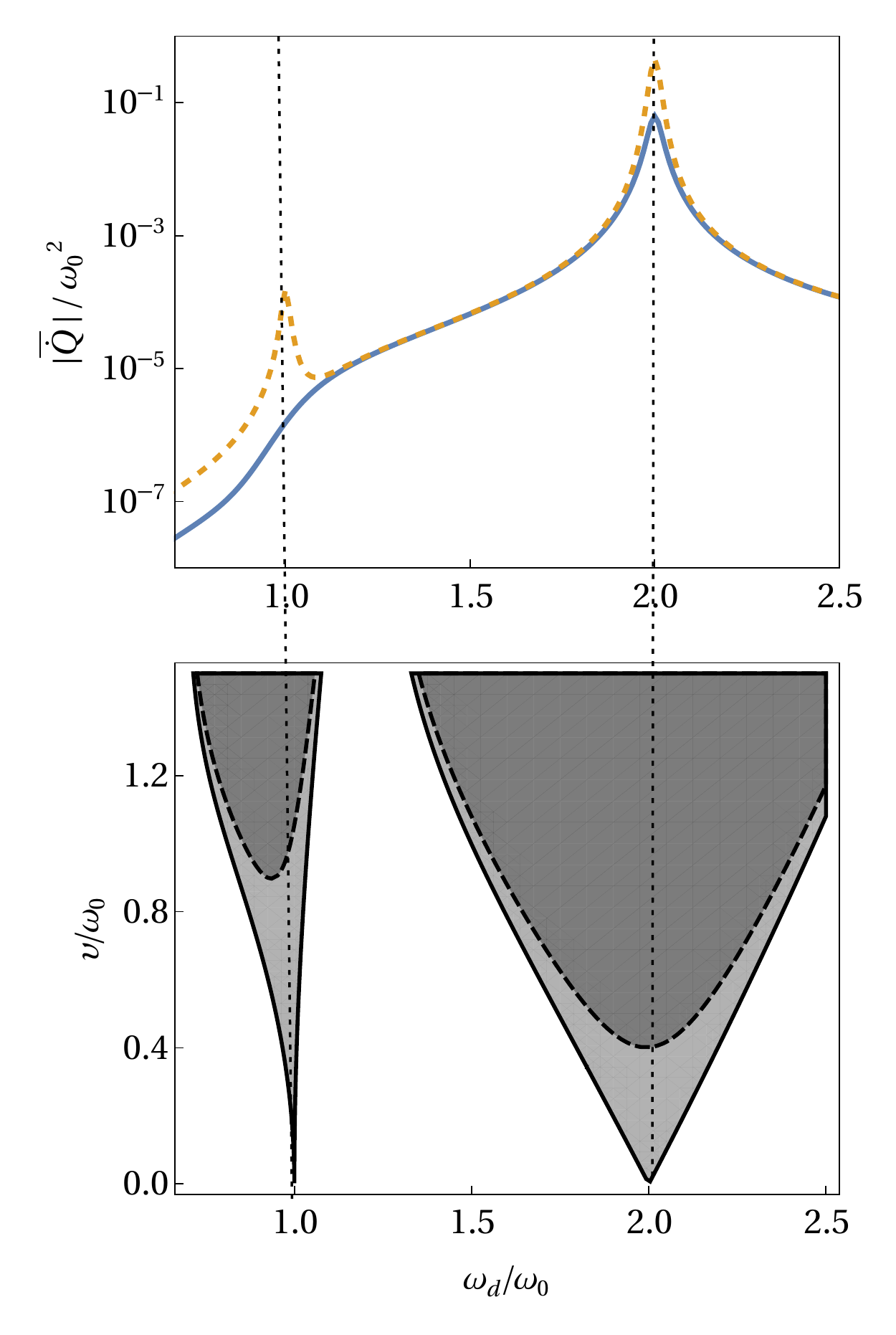}
\caption{\textbf{(top)} Cycle-averaged heating rate of the sample as a function of the driving frequency $\omega_d$. The heat current is calculated to second and fourth order in $\upsilon$ (solid blue and dashed orange, respectively). As we can see, the heating around resonance is a fourth-order effect. All parameters are as in Fig.~\ref{fig1} and $ T=0.025 $. \textbf{(bottom)} Stability chart of a (classical) damped Mathieu oscillator for varying driving frequency $ \omega_d $ and driving amplitude $\upsilon$, at fixed $ \omega_0 $. Unstable solutions of exponentially growing amplitude appear in the shaded regions. In absence of friction ($\gamma=0$; light grey) solutions are unstable at the resonance frequencies $ \omega_d = 2\omega_0/n $ for integer $ n $ (vertical dotted lines), regardless of $ \upsilon $. On the contrary, stable solutions may always be found at sufficiently weak driving under non-zero friction. For instance, unstable solutions for $ \gamma=0.1 $ are limited to the dark grey area.}
\label{fig4}
\centering
\end{figure}

So far we have seen how periodic driving can substantially boost the sensitivity of low-temperature thermometry. This, however, comes at the cost of heating the sample as the work pumped into the probe is dissipated away. In order to assess the magnitude of this effect, we now focus on characterising the probe--sample energy exchanges in the limit cycle. 

The rate of change of the energy of the probe is simply $ \frac{d}{dt}\langle\pmb H_S(t)\rangle = - i \langle [ \pmb H_S(t), \pmb H_I] \rangle + \langle \partial_t\,\pmb H_S(t) \rangle $, where the brackets indicate averaging over the instantaneous state. Specifically, the first term is the contribution due to the probe--sample coupling. Let us denote it by
\begin{equation}\label{eq:instantaneous-heat-current}
	\dot{Q}(t) = - i \langle [ \pmb H_S(t), \pmb H_I] \rangle.
\end{equation}
In the limit cycle, this term may be cast as (see Appendix \ref{app:current})
\begin{equation}
	\dot{Q}(t) = \frac{1}{2} \frac{d}{dt}\sigma_{pp}(t)  +\left(\omega^2(t)+\omega_R^2 \right)\sigma_{xp}(t).
\end{equation}
Energy conservation and stationarity require that the asymptotic cycle average of $\dot{Q}(t)$, i.e., 
\begin{equation*}
	\overline{\dot{Q}} = \frac{\omega_d}{2 \pi} \int_0^{2 \pi/\omega_d} \dot{Q}^\text{lc}(t)\,dt,
\end{equation*}
coincides with minus the average heating rate of the sample \cite{freitas2017linear_refrigerator}. Explicitly, $ \overline{\dot{Q}} $ takes the form \begin{multline}
	\overline{\dot{Q}}=-\frac{1}{\pi} \int_0^\infty \sum\nolimits_k\,k\,\omega_d\, \tilde{J}(\omega+k\,\omega_d)\,a_k(\omega)\,a^*_k(\omega)\\\times J(\omega)\,\coth{(\omega/2T)}\, d\omega, \label{curr2}
\end{multline}
where $ \tilde{J}(\omega) $ is the spectral density extended to negative frequencies as an odd function (cf. Appendix~\ref{app:current}). 

Fig.~\ref{fig4} shows the limit-cycle-averaged heat current into the sample as a function of the driving frequency. It has been calculated both to second and fourth order in $\upsilon$, assuming again the Ohmic spectral density in Eq.~\eqref{eq:spectral_density}.

We see a large peak in the heat current around $\omega_d=2 \omega_0$, and a smaller one (a fourth-order effect) at $\omega_d=\omega_0$. Tuning the drive close to resonance to enhance sensitivity thus comes at the cost of a larger heating of the sample. The compromise between the two will be ultimately dictated by the heat capacity of the sample. This is, however, typically very large, so that fourth-order heating effects would be negligible at small driving amplitude $ \upsilon $. When it comes to the rest of parameters in the problem, smaller dissipation strength and lower temperature both reduce the heat current.

The dependence of the current on $\omega_d$ suggests a connection with parametric resonance. Indeed, studying the closely related classical Mathieu oscillator we find some striking similarities. The damped Mathieu equation reads
\begin{equation}
	\ddot{x} + \gamma\,\dot{x} + \omega^2(t)\,x=0. \label{eq:Mathieu_osc}
\end{equation}

In the limit of $\omega_c \rightarrow \infty$ the integral in Eq.~\eqref{eq:QLE_driven} generates a dissipation term of the form $\gamma \dot{x}$. The main difference between the Mathieu equation and our model (besides the former being classical) is the absence of thermal noise in Eq.~\eqref{eq:Mathieu_osc}. 

The solutions of the Mathieu equation as time evolves are either decaying, purely oscillatory, or have increasing amplitude (see Appendix \ref{app:mathieu}). The latter are termed unstable. Fig.~\ref{fig4} shows the regions of instability of the solution to Eq.~\eqref{eq:Mathieu_osc} in parameter space. In the absence of damping (i.e., $\gamma=0$) we recover the well-known instabilities at parametric resonance. Specifically, at $\omega_d= 2\omega_0/n$ for $n \in \{1,2,...\}$. Those are precisely the points at which the heat current peaks in our model, thus suggesting a connection between heating and instability. We do work, however, at finite dissipation strength and small $ \upsilon $, which would allow to avoid instabilities even near resonance.

\section{Conclusions}\label{sec:conclusions}

We have studied how the precision of low-temperature measurements can be largely improved by periodic driving. More concretely, we have solved for the long-time limit of a driven quantum Brownian temperature probe in an equilibrium sample. We have found that the low-temperature sensitivity of the Brownian probe may be substantially boosted by simply applying a weak sinusoidal near-resonant modulation to its harmonic confining potential. 

In order to quantify thermal sensitivity, we have calculated the upper limit on the signal-to-noise ratio $ \frac{1}{\sqrt{N}}\,(T/\delta T_\text{est}) $. For a large number of measurements $ N $, this is given by the quantum Fisher information of the limit-cycle probe state. Interestingly, we have shown that the boost in precision does survive limit-cycle averaging. In other words, enabling a precision improvement of several orders of magnitude in a thermometry experiment would not require synchronisation between the measurements and the drive. 

Furthermore, we have illustrated how the exponent of the power-law-like asymptotic behaviour of the signal-to-noise ratio at low $T$ \cite{hovhannisyan2018termometry_many-body} may be manipulated by the driving. Effectively, this renders low-temperature thermometry much more efficient than otherwise allowed by the spectral density of the sample. Specifically, we show that the ultimate precision limit can switch from a $ \sim T^2 $ scaling to a $ \sim T $ behaviour as $ T\rightarrow 0 $ on a sample with Ohmic spectrum. 

Finally, we have exploited the fact that an impurity thermometry setup in a BEC can be microscopically modelled with our linear quantum Brownian motion setup. We have shown that, for representative experimental parameters, weak near-resonant driving of the impurity trap does lead to a very substantial amplification of thermal sensitivity for $ T\lesssim \SI{1}{\nano\kelvin} $. In this case, we used a practically motivated proxy for thermal sensitivity---the responsiveness of the time-averaged density profile of the impurity gas to temperature fluctuations. 

We have also investigated the impact of heat dissipation onto the sample as a byproduct of the driving. We have found that, when driving near resonance, heating is a fourth-order effect in the driving strength, thus producing minimal back-action on the ultracold sample.

Crucially, the explicit formulae that we have derived for the asymptotic state of the probe are \textit{exact} to any order in the strength of the external drive. Since we have avoided using Markovian quantum master equations, we have bypassed the severely limiting assumptions of fast-decaying bath correlation functions and weak system--bath coupling. Although the correct intuition can be drawn from a simple Markovian calculation \cite{mukherjee2019thermometry_control}, our results extend to arbitrary coupling and spectral density, and to arbitrarily low temperatures, thus permitting the study of the $ T\rightarrow 0 $ scaling of thermometric precision. 

Ultimately, the physics behind the sensitivity improvement is the closing of the Floquet quasi-energy gap, which has been exploited in other sensing applications (see, e.g., \cite{PhysRevLett.127.080504}). We thus expect our results to qualitatively hold beyond our particular model. Indeed, in the limit of weak dissipation, periodic driving can be seen to have the same effect in non-harmonic probes \cite{mukherjee2019thermometry_control,alicki2012periodically}. Similarly, since the most crucial role for the question at hand is played by the low-frequency tail of the spectral density, we also expect our results to extend beyond the Ohmic and super-Ohmic spectral densities considered here.

We have thus validated a robust solution to increase low-temperature sensitivity in thermometry experiments, and quantified its impact on thermometry in ultracold atomic gases. Studying the potential benefits on other platforms and temperature ranges---e.g., $ \SI{}{\milli\kelvin} $-cooled superconducting circuits---remains an interesting open direction. Likewise, it would be practically relevant to assess the impact of periodic driving when working with small measurement records on both the accuracy and the speed of convergence of a temperature estimate. This may be achieved with the newly developed global (Bayesian) thermometry tools \cite{rubio2021global} and will be the subject of future work.

\acknowledgements

We thank M. Mehboudi, A. Riera-Campeny, N. Freitas, P. Potts, J. Anders, M. Perarnau-Llobet and G. Barontini for useful discussions. JG is supported by a scholarship from the College of Engineering, Mathematics and Physical Sciences of the University of Exeter.

\appendix

\renewcommand{\thefootnote}{\fnsymbol{footnote}}

\section{Limit cycle of the probe}\label{app:solving}

\subsection{Equation of motion}\label{app:eom}
The Heisenberg equations of motion from Eqs.~(\ref{eq:system_Hamiltonian}--\ref{eq:sys-bath}) are
\begin{subequations}
\begin{align}
	\dot{\pmb x} &= \pmb p,\\
	\dot{\pmb p} &= - \big(\omega(t)^2+\omega_R^2\big) \pmb x - \sum\nolimits_\mu g_\mu\,\pmb x_\mu,\label{eq:dpdt}\\
	\dot{\pmb x}_\mu &= \pmb p_\mu/m_\mu,\\
	\dot{\pmb p}_\mu &= - m_\mu \omega_\mu^2\,\pmb x_\mu - g_\mu\,\pmb x.
\end{align}
\end{subequations}
To make this paper self-contained, we shall now provide full detail on how to obtain their limit cycle, following Ref.~\cite{freitas2017linear_refrigerator}. To simplify the notation we define $ \pmb{\mathsf{Z}}_\mu = (\pmb x_\mu, \pmb p_\mu)^\mathsf{T}$ and $ \pmb{\mathsf{Z}} = (\pmb x, \pmb p)^\mathsf{T} $. Hence, dynamical equations for the sample degrees of freedom can be compacted as
\begin{equation}\label{eq:Heisenberg-sample}
	\dot{\pmb{\mathsf{Z}}}_\mu(t) + \begin{pmatrix}0 & - m_\mu^{-1}\\ m_\mu \omega_\mu^2&0\end{pmatrix}\, \pmb{\mathsf{Z}}_\mu(t)  = \begin{pmatrix}0 & 0\\ -g_\mu&0\end{pmatrix}\, \pmb{\mathsf{Z}}(t),
\end{equation}
where $\pmb{\mathsf{Z}}(t)$ acts as a source term in the equations for $\pmb{\mathsf{Z}}_\mu(t)$. The corresponding Green's function $ \mathsf{G}_\mu (t,t') $ thus satisfies
\begin{multline*}
	\partial_t\, \mathsf{G}_\mu(t,t') + \begin{pmatrix}0 & -m_\mu^{-1}\\ m_\mu \omega_\mu^2&0\end{pmatrix} \mathsf{G}_\mu(t,t')  \\ = \delta(t-t')\,\mathbbm{1}_2,
\end{multline*}
where $\mathbbm{1}_2$ is the $2\times 2$ identity matrix and $ \delta(x) $ stands for the Dirac delta. Hence,
\begin{multline*}
    \mathsf{G}_\mu(t,t')  =  
	\Theta(t-t') \\ \times\begin{pmatrix}\cos{(\omega_\mu (t-t'))} & \frac{\sin{(\omega_\mu (t-t'))}}{m_\mu\,\omega_\mu} \\-m_\mu \omega_\mu \sin{(\omega_\mu (t-t'))}&\cos{(\omega_\mu (t-t'))}\end{pmatrix}.
\end{multline*}
Therefore, Eq.~\eqref{eq:Heisenberg-sample} is then solved by
\begin{equation}
	\pmb{\mathsf{Z}}_\mu(t) = \mathsf{G}_\mu(t,0)\, \pmb{\mathsf{Z}}_\mu(0) + \int_0^t \mathsf{G}_\mu(t,t')\,\mathsf{C}_\mu\,\pmb{Z}(t')\,dt',
\end{equation}
where $\mathsf{C}_\mu = \begin{pmatrix}0 & 0\\ -g_\mu&0\end{pmatrix}$.

Similarly, we can now look at the equations of motion of the degrees of freedom of the probe
\begin{multline}
	\dot{\pmb{\mathsf{Z}}}(t) + \begin{pmatrix}0 & -1\\ \omega(t)^2 + \omega_R^2 &0\end{pmatrix}  \pmb{\mathsf{Z}}(t) \\- \int_0^t  \begin{pmatrix}0 & 0\\\chi(t-t')&0\end{pmatrix} \pmb{\mathsf{Z}}(t') dt'  = \pmb{\mathsf{F}}(t) \label{eq:vectorised-eom-probe},
\end{multline}
where the dissipation kernel $\chi(t-t')$ was introduced in Eq.~\eqref{eq:dissipation_kernel} in the main text, and $\pmb{\mathsf{F}}(t) = \sum_\mu \mathsf{C}_\mu\,\mathsf{G}_\mu(t,0)\,\pmb{\mathsf{Z}}_\mu(0)$. Note that Eq.~\eqref{eq:vectorised-eom-probe} can be recast as a second-order equation for $\pmb x$ of the form advanced in Eq.~\eqref{eq:QLE_driven} of the main text, with stochastic force $\pmb{F}(t)$ given by
\begin{multline*}
	\pmb F(t) = - \sum_\mu g_\mu\, \\ \times\left( \pmb x_\mu(0) \cos{(\omega_\mu t)} + \frac{\pmb p_\mu(0)}{m_\mu \omega_\mu} \sin{(\omega_\mu t)} \right).
\end{multline*}
Back to Eq.~\eqref{eq:vectorised-eom-probe}, the corresponding Green's function $\mathsf{G}(t,t')$ satisfies
\begin{multline}\label{eq:Green-function-eq-system}
	\partial t\,\mathsf{G}(t,t') + \begin{pmatrix}0 & -1\\ \omega(t)^2 + \omega_R^2 &0\end{pmatrix}  \mathsf{G}(t,t') \\
	- \int_0^t  \begin{pmatrix}0 & 0\\ \chi(t-\tau)&0\end{pmatrix} \mathsf{G}(\tau,t') \, d \tau = \delta(t-t')\,\mathbbm{1}_2.
\end{multline}
The equation has the formal solution
\begin{equation}
	\pmb{\mathsf{Z}}(t) = \mathsf{G}(t,0)\,\pmb{\mathsf{Z}}(0) + \int_0^t \mathsf{G}(t,t')\,\pmb{\mathsf{F}}(t')\,dt' \label{ansatz},
\end{equation}
which will allow us to find $ \mathsf{G}(t,t') $ in the limit cycle.

\subsection{Covariance matrix in the limit cycle} \label{app:covcycle}

The covariance matrix takes the form
\begin{multline*}
	\mathsf{\Sigma}(t) = \textrm{Re}\,\langle \pmb{\mathsf{Z}}(t)\pmb{\mathsf{Z}}(t)^\mathsf{T} \rangle - \langle \pmb{\mathsf{Z}}(t)\rangle \langle \pmb{\mathsf{Z}}(t)^\mathsf{T}\rangle \\= \begin{pmatrix} \sigma_{xx}(t) & \sigma_{xp}(t) \\ \sigma_{xp}(t) & \sigma_{pp}(t) \end{pmatrix}.
\end{multline*}
For initial states with $\langle \pmb{\mathsf{Z}}(0)\rangle = \langle \pmb{\mathsf{Z}}_\mu(0)\rangle = 0$, the first moments vanish for all times, as can be seen from \eqref{ansatz}. Since $\langle \pmb{\mathsf{Z}}_\mu(0)\rangle=0$ is automatically fulfilled for a sample in a thermal state, we only need to assume $\langle \pmb{\mathsf{Z}}(0)\rangle = 0$.

From Eq.~\eqref{ansatz} the time evolution of the covariance matrix can be written as
\begin{widetext}
\begin{equation*}
\mathsf{\Sigma}(t) = \mathsf{G}(t,0)\,\mathsf{\Sigma}(0)\,\mathsf{G}(t,0)^\mathsf{T} + \mathsf{G}(t,0)\,\textrm{Re}\,\langle \pmb{\mathsf{Z}}(0)\pmb{\mathsf{B}}(t)^\mathsf{T} \rangle
	+ \textrm{Re}\,\langle \pmb{\mathsf{B}}(t)\pmb{\mathsf{Z}}(0)^\mathsf{T} \rangle \,\mathsf{G}(t,0)^\mathsf{T}+\textrm{Re}\,\langle\pmb{\mathsf{B}}(t)\pmb{\mathsf{B}}(t)^\mathsf{T} \rangle,
\end{equation*}
\end{widetext}
with $\pmb{\mathsf{B}}(t)= \int_0^\infty\,\mathsf{G}(t,t')\,\pmb{\mathsf{F}}(t')\,dt'$. The second and third term are propagations of initial probe--sample correlations and therefore vanish in our case. In turn, the first term decays for $t \rightarrow \infty$ for exponentially stable systems. Here, we assume that this is indeed the case and later discuss when this assumption fails; namely, at parametric resonance (see section \ref{sec:heating} and Appendix \ref{app:mathieu}). We are thus left with
\begin{widetext}
\begin{align}\label{eq:limit-cycle-covariance-raw}
	\mathsf{\Sigma}^\text{lc}(t) = \textrm{Re}\, \langle \pmb{\mathsf{B}}(t)\pmb{\mathsf{B}}(t)^\mathsf{T}\rangle = \int_0^t \int_0^t \mathsf{G}(t,t_1) \sum\nolimits_{\mu,\nu} \mathsf{C}_\mu \mathsf{G}_\mu(t_1,0)\,\textrm{Re}\, \langle \pmb{\mathsf{Z}}_\mu(0)\pmb{\mathsf{Z}}_\nu(0)^\mathsf{T} \rangle \mathsf{G}_\nu(t_2,0)^\mathsf{T} \mathsf{C}^\mathsf{T}_\nu \mathsf{G}(t,t_2)^\mathsf{T}\,dt_1\,dt_2 ,
\end{align}
\end{widetext}
where, as stated in the main text, the superscript `lc' stands for `limit cycle'. Since the sample is in thermal equilibrium at temperatures $T$
\begin{multline*}
	\textrm{Re}\,\langle \pmb{\mathsf{Z}}_\mu(0)\,\pmb{\mathsf{Z}}_\nu(0)^\mathsf{T} \rangle \\= \delta_{\mu,\nu} \begin{pmatrix} \frac{1}{2\,m_\mu \omega_\mu} \coth\frac{\omega_\mu}{2 T}&0\\0 &\frac{m_\mu \omega_\mu}{2} \coth\frac{\omega_\mu}{2 T}\end{pmatrix},
\end{multline*}
where $\delta_{\mu,\nu}$ is a Kronecker delta. Eq.~\eqref{eq:limit-cycle-covariance-raw} thus simplifies to
\begin{multline}\label{eq:covariance-limit-cycle_general}
\mathsf{\Sigma}^\text{lc}(t) = \frac12\,\int_0^t \int_0^t \mathsf{G}(t,t_1)\\ \times\begin{pmatrix} 0&0\\0 &\mu(t_1-t_2)\end{pmatrix} \mathsf{G}(t,t_2)^\mathsf{T} dt_1 dt_2,
\end{multline}
with the ‘noise kernel’ $ \mu(t) $ given by
\begin{align}\label{eq:noise-kernel}
\mu(t) = \frac{2}{\pi} \Theta(t) \int_0^\infty J(\omega)\,\cos{\omega t}\, \coth{\frac{\omega}{2 T}}\,d\omega.
\end{align}

Eq.~\eqref{eq:covariance-limit-cycle_general} reduces to the following expressions for the elements of the covariance matrix
\begin{subequations}\label{eq:limit-cycle-covariances-time}
\begin{align}
	\sigma_{xx}^\text{lc}(t) &= \frac{1}{2}  \int_0^t \int_0^t g(t,t_1)\mu(t_1-t_2) \nonumber\\ &\quad\qquad\qquad\qquad \times g(t,t_2)\,dt_1\,dt_2, \\
	\sigma_{xp}^\text{lc}(t) &= \frac{1}{2}  \int_0^t \int_0^t g(t,t_1)\mu(t_1-t_2) \nonumber\\ &\quad\qquad\qquad\qquad\times\partial_t g(t,t_2)\,dt_1\,dt_2, \\
	\sigma_{pp}^\text{lc}(t) &= \frac{1}{2}  \int_0^t \int_0^t \partial_t\,g(t,t_1)\mu(t_1-t_2) \nonumber \\ &\quad\qquad\qquad\qquad\times\partial_t g(t,t_2)\,dt_1\,dt_2,
\end{align}
\end{subequations}
where we have introduced the notation $ g(t,t') \coloneqq  [\mathsf{G}(t,t')]_{12} $, and used $[\mathsf{G}(t,t')]_{22} = \partial_t [\mathsf{G}(t,t')]_{12}$, which follows from Eq.~\eqref{eq:Green-function-eq-system}. Exploiting this identity, one further sees that
\begin{multline}\label{eq:Langevin_g}
    \partial_t^2 g(t,t') + \big(\omega^2(t)+\omega_R^2\big)\, g(t,t') \\- \int_0^t \chi(t-\tau)\,g(\tau,t')\,d\tau = \delta(t-t'),
\end{multline}
that is, $ g(t,t') $ is the Green's function of the equation of motion  \eqref{eq:QLE_driven} for the system's $ \pmb x(t) $.

Eqs.~\eqref{eq:limit-cycle-covariances-time} can be manipulated further by casting the noise kernel \eqref{eq:noise-kernel} as 
\begin{equation*}
    \mu(t) = \textrm{Re}\,\int_0^\infty \hat{\mu}(\omega)\,\mathrm{e}^{i\omega t}\,d\omega,
\end{equation*} 
with $ \hat{\mu}(\omega) \coloneqq \frac2\pi\,J(\omega)\,\coth{(\omega/2T)} $, as defined in the main text. This gives 
\begin{subequations}\label{eq:limit-cycle-covariances-general}
\begin{align}
	\sigma_{xx}^\text{lc}(t) &=\frac{1}{2}\, \textrm{Re}\,\int_0^\infty  q(t,\omega)\,\hat{\mu} (\omega)\,q(t,\omega)^* d\omega, \label{asxx} \\
	\sigma_{xp}^\text{lc}(t) &=\frac{1}{2}\, \textrm{Re}\, \int_0^\infty q(t,\omega)\,\hat{\mu} (\omega)\,\partial_t q(t,\omega)^* d\omega, \\
	\sigma_{pp}^\text{lc}(t) &=\frac{1}{2}\, \textrm{Re}\, \int_0^\infty  \partial_t q(t,\omega)\,\hat{\mu} (\omega)\, \partial_t q(t,\omega)^* d\omega, \label{aspp}
\end{align}
\end{subequations}
where $ q(t,\omega) = \int_0^t g(t,t')\,\mathrm{e}^{i \omega t'}\,dt' $.

Recalling that our probe is periodically driven, i.e., $ \pmb H_S(t+\tau) = \pmb H_S(t)$ with $\tau = 2 \pi/\omega_d$, we must have $g(t+\tau,t'+\tau)=g(t,t')$. Therefore,
\begin{align*}
	q(t+\tau,\omega)&=\int_0^{t+\tau}  g(t+\tau,t')\,\mathrm{e}^{i \omega t'}\,dt'\nonumber \\
	&= \int_{-\tau}^{t} g(t+\tau,t'+\tau)\,\mathrm{e}^{i \omega (t'+\tau)}\,dt' \nonumber \\
	&= \mathrm{e}^{i \omega \tau}\,\int_{-\tau}^{t} g(t,t')\,\mathrm{e}^{i \omega t'}\,dt'.
\end{align*}
The probe reaching a limit cycle requires $g(t,t')$ going to zero sufficiently fast for $\abs{t-t'} \gg 1$. In the limit cycle, i.e, $t \gg \tau$, we therefore have
\begin{equation*}
	\int_{-\tau}^{t} g(t,t')\,\mathrm{e}^{i \omega t'}\, dt' \simeq \int_0^{t} g(t,t')\,\mathrm{e}^{i \omega t'}\,dt' = q(t,\omega).
\end{equation*}
That is, $q(t+\tau,\omega) = \mathrm{e}^{i\omega\tau}\,q(t,\omega)$. Hence, the function $p(t,\omega) \coloneqq \mathrm{e}^{-i\omega t}\,q(t,\omega)$ is $\tau$-periodic in the limit cycle. As a result, we may write its Fourier series as 
\begin{multline}\label{eq:p_Fourier_series}
	p(t,\omega) =\sum_{k=-\infty}^{\infty} a_k (\omega)\,\mathrm{e}^{i k \omega_d t} \\= \mathrm{e}^{-i\omega t}\int_0^t g(t,t') \mathrm{e}^{i\omega t'}\,dt'.
\end{multline}
Using $ \delta(t) = \frac{1}{2\pi}\int_{-\infty}^{\infty} \mathrm{e}^{i\omega t} d\omega $ it is easy to see that 
\begin{multline}\label{eq:p(t,w)_fourier_series}
    g(t,t') = \frac{1}{2\pi}\int_{-\infty}^{\infty} p(t,\omega)\,\mathrm{e}^{i\omega(t-t')}\,d\omega \\
    = \frac{1}{2\pi}\sum_{k=-\infty}^{\infty}\int_{-\infty}^{\infty} a_k(\omega)\,\mathrm{e}^{i\omega(t-t')}\,\mathrm{e}^{ik\omega_d t}\,d\omega,
\end{multline}
which is Eq.~\eqref{eq:Green-function-and-As} from the main text. 

Inserting now \eqref{eq:p(t,w)_fourier_series} into \eqref{eq:Langevin_g} gives
\begin{multline*}
    \frac{1}{2\pi}\int_{-\infty}^\infty d\omega\,\mathrm{e}^{i\omega(t-t')} \\ \times\sum\nolimits_k \mathrm{e}^{ik\omega_d t}\big[ a_k(\omega)\,\hat{g}_0[i(\omega+k\omega_d)]^{-1} \\+ \sum\nolimits_{l\neq 0} a_{k-l}(\omega)b_l\big] = \delta(t-t').
\end{multline*}
Using again $ \delta(t-t') = \frac{1}{2\pi}\int_{-\infty}^{\infty} \mathrm{e}^{i\omega (t-t')} d\omega $ readily brings us to Eq.~\eqref{eq:A} in the main text, i.e.,
\begin{equation}\label{eq:ak-bl-equation}
    \hat{g}_0[i (\omega + k \omega_d)]^{-1}a_k(\omega) + \sum\nolimits_{l \neq 0}\,b_l\,a_{k-l} (\omega) = \delta_{k0},
\end{equation}
where the frequency has been expanded as
\begin{equation*}
    \omega^2(t) = \omega_0^2 + \sum\nolimits_{l\neq 0 } b_l\,\mathrm{e}^{i\,l\,\omega_d\,t}.
\end{equation*}

Finally, writing $ q(t,\omega) = e^{i\omega t} p(t,\omega) $ and using the Fourier series \eqref{eq:p_Fourier_series} allows to rewrite Eqs.~\eqref{eq:limit-cycle-covariances-general} as Eqs.~\eqref{eq:covariances-limit-cycle} and \eqref{eq:covariances_coefficients-limit-cycle} in the main text.

\section{Heating of the sample}\label{app:current}

The rate of change of the energy of the probe can be expressed as
\begin{align}\label{eq:energy-change-explicit}
    \frac{d}{d t} \langle \pmb H_s(t) \rangle &= - i \langle \left[ \pmb H_S(t),\pmb H(t) \right] \rangle +  \partial_t \langle \pmb H_S(t) \rangle \nonumber \\
    &=- i \langle \left[ \pmb H_S(t), \pmb H(t) \right] \rangle + \frac{1}{2} \partial_t \omega(t)^2\,\sigma_{xx}(t).
\end{align}
Here, the last term is the input power from the external modulation of the trapping frequency. Hence, we write
\begin{align*}
	\dot{W}(t) &\coloneqq\frac{1}{2} \partial_t \omega(t)^2\,\sigma_{xx}(t).
\end{align*}
In turn, as we saw in Sec.~\ref{sec:heating} above, the remaining term is the change of energy due to the the probe--sample interaction, denoted $ \dot{Q}(t) $. Evaluating the commutator in \eqref{eq:instantaneous-heat-current} gives
\begin{align*}
    \dot{Q}(t)&= - \langle \textstyle\sum\nolimits_\mu g_\mu\,\pmb x_\mu\,\pmb p \rangle = -\frac12\langle\{ \textstyle\sum_\mu g_\mu \pmb x_\mu,\pmb p \}\rangle.
\end{align*}
Using now the equation of motion \eqref{eq:dpdt} allows us to write
\begin{align}\label{eq:Qt_step}
    \dot{Q}(t) &= \frac12\langle\{\dot{\pmb p},\pmb p\}\rangle + (\omega(t)^2 + \omega_R^2)\,\frac12\langle\{\pmb x, \pmb p\}\rangle \nonumber\\
    &= \frac{1}{2} \frac{d}{d t} \sigma_{pp}(t) + (\omega(t)^2+\omega_R^2)\, \sigma_{xp}(t).
\end{align}

We now place ourselves at the limit cycle, thus working with $ \sigma_{pp}^\text{lc}(t) $ and $ \sigma_{xp}^\text{lc}(t) $, and carry out the cycle average
\begin{equation*}
    \overline{\dot{Q}} \coloneqq \frac{2\pi}{\omega_d}\int_0^{2\pi/\omega_d} \dot{Q}^\text{lc}(t)\,dt.
\end{equation*}
Note that the contribution to $ \overline{\dot{Q}} $ from the first term on the right-hand side of \eqref{eq:Qt_step} would then vanish, as per Eq.~\eqref{eq:sxxt}. In turn, using Eqs.~\eqref{eq:frequency-Fourier} and \eqref{eq:sxpt} leads to
\begin{multline*}
	\overline{\dot{Q}} =\frac{1}{2} \im{\int_0^\infty \sum\nolimits_{j,k} b_{k-j}\, (\omega + k \omega_d) \\ \times a_j(\omega)\,\hat{\mu}(\omega)\,a_k^*(\omega)\,d\omega},
\end{multline*}
where $ b_0 = \omega_0^2+\omega_R^2 $. We now note that
\begin{equation*}
\im{\sum\nolimits_{j,k} b_{k-j} \,a_j(\omega)\,a_k^*(\omega)} = 0,
\end{equation*}
since $ b_{j-k} = {(b_{k-j})}^* $. Therefore, given that $ \hat{\mu}(\omega) $ is also real, we may write the simplified expression
\begin{equation*}
	\overline{\dot{Q}} =\frac{\omega_d}{2}\,\im{\int_0^\infty \sum\nolimits_{j,k} k \,b_{k-j}\, a_j(\omega)\,\hat{\mu}(\omega)\,a_k^*(\omega)\,d\omega}.
\end{equation*}
Using equation \eqref{eq:ak-bl-equation} we can write
\begin{equation*}
	\sum\nolimits_j b_{k-j}\,a_j(\omega) = \delta_{k0}-(\hat{g}_0[i(\omega + k \omega_d)]^{-1}-b_0)\,a_k(\omega),
\end{equation*}
so that the limit-cycle-averaged heat current becomes
\begin{multline*}
	\overline{\dot{Q}} =-\frac{\omega_d}{2}\,\im{ \int_0^\infty \sum\nolimits_{k} k\,  (\hat{g}_0[i(\omega + k \omega_d)]^{-1}-b_0)\\\times a_k(\omega)\,\hat{\mu}(\omega)\,a_k^*(\omega)\,d\omega }.
\end{multline*}
To complete the derivation we use the identity \cite{correa2017thermometry_strongcoupling}
\begin{multline*}
    \im{(\hat{g}_0(i\omega)^{-1}-b_0)}= \im{\hat{\chi}(i\omega)} \\= J(\omega)\,\Theta(\omega)-J(-\omega)\,\Theta(-\omega)\coloneqq \tilde{J}(\omega).
\end{multline*}
Writing $\hat{\mu}(\omega)$ explicitly as per Eq.~\eqref{eq:nk} finally gives us Eq.~\eqref{curr2} from the main text; namely, 
\begin{multline}
	\dot{\bar{Q}}=-\frac{1}{\pi} \int_0^\infty \sum_k k \omega_d  J(\omega+k\omega_d) a_k(\omega) \\ \times J(\omega) a^*_k(\omega) \coth{\frac{\omega}{2T}}\,d\omega.
\end{multline}

\section{Mathieu equation and parametric resonance}\label{app:mathieu}

As stated above, the damped Mathieu equation reads
\begin{equation*}
	\ddot{x} + \gamma\,\dot{x} + \omega^2(t)\,x = 0,
\end{equation*}
with $ \omega^2(t)=\omega_0^2+\upsilon \cos{\omega_d t} $. Introducing the dimensionless parameters $\tilde{t}=\omega_d\,t/2$, $\tilde{\omega}_0^2=4 \omega_0^2/\omega_d^2$, $\tilde{\upsilon}=2\upsilon/\omega_d^2$, and $\tilde{\gamma}=2\gamma/\omega_d$, the equation can be recast as
\begin{equation*}
	\ddot{x}+\tilde{\gamma}\,\dot{x} + \tilde{\omega}^2(\tilde{t})\,x = 0,
\end{equation*}
with $ \tilde{\omega}^2(t) = \tilde{\omega}_0^2+2\,\tilde{\upsilon}\,\cos{2 \tilde{t}} $. Applying now the transformation $ x = y\,\mathrm{e}^{-\tilde{\gamma}\,\tilde{t}/2}$ allows to absorb the damping into a frequency shift, bringing the Mathieu equation into the form
\begin{equation*}
	\ddot{y} + ( \tilde{\omega}_0^2-\tilde{\gamma}^2/4+2\,\tilde{\upsilon}\,\cos{2\, \tilde{t}} )\,y = 0.
\end{equation*}
This is solved by a function of the form \cite{zerbe1995parametric,nayfeh1995}
\begin{equation*}
	y(\tilde{t}) = c_1\,\mathrm{e}^{i\,\nu\,\tilde{t}}\,p(\tilde{t}) + c_2\,\mathrm{e}^{- i\,\nu\,\tilde{t}}\,p(-\tilde{t}),
\end{equation*}
where $\nu$ is the characteristic exponent, whose dependence on the parameters $\tilde{\omega}_0$, $\tilde{\gamma}$, and $\tilde{v}$ can be readily explored numerically. The functions $p(\tilde{t})$ here are periodic. Undoing the previous change of variables back to $ x $ then gives
\begin{equation*}
	x(\tilde{t}) = c_1\,\mathrm{e}^{(i\,\nu - \tilde{\gamma}/2 )\,\tilde{t}} p(\tilde{t}) + c_2\,\mathrm{e}^{- (i\,\nu - \tilde{\gamma}/2)\,\tilde{t}} p(-\tilde{t}).
\end{equation*}
The asymptotic behaviour of this solution depends on $\im{\nu}$; namely, if $\im{\nu} > \tilde{\gamma}/2$ the amplitude grows exponentially; if $ \im{\nu} < \tilde{\gamma}/2$ the amplitude decays to zero; and if $\im{\nu} = \tilde{\gamma}/2$, the solution is purely oscillatory.

\bibliography{references}

\end{document}